\documentclass[aps,prb,twocolumn,superscriptaddress,floatfix]
{revtex4-1}
\usepackage{graphicx}
\usepackage{amsmath,amssymb}
\begin{document}
\title{Non-Gaussian diffusion profiles caused by mobile impurity-vacancy
pairs in the five frequency model of diffusion}
\author{V. I. Tokar}
\affiliation{IPCMS, Universit{\'e} de Strasbourg--CNRS, UMR 7504,
23 rue du Loess, F-67034 Strasbourg, France}
\date{\today}
\begin{abstract}
Vacancy-mediated diffusion of impurities under strong impurity-vacancy
(I-v) attraction has been studied in the framework of the five-frequency
model (5FM) for the FCC host.  The system of impurities and tightly
bound I-v pairs has been treated in the framework of the rate-equations
approach of Cowern et al., Phys. Rev.  Lett. {\bf 65}, 2434 (1990),
developed for the description of the non-Gaussian diffusion profiles
(NGDPs) observed in dopant diffusion in silicon. In the present study
this approach has been extended to derive a three-dimensional (3D)
integro-differential equation describing the pair-mediated impurity
diffusion. The equation predicts the same 1D NGDPs as in Cowern et al.\
but can be also used for the simulation of 3D profiles of arbitrary
geometry in the systems where the diffusion proceeds via a mobile state.
The parameters of the theory has been calculated within the 5FM on
the basis of available literature data.  The database on impurities in
aluminum host has been analyzed and promising impurity-host systems for
the observation of NGDPs has been identified.  The diffusion profiles for
an impurity where NGDPs are expected to be easily detectable have been
simulated. It has been argued that with the input parameters calculated
on the basis of experimental diffusion constants the simulated NGDPs
can be accurate enough to serve as a quantitative test of the 5FM.
\end{abstract}
\maketitle 
\section{Introduction}
Miniaturization of electronic devices to nanometer sizes has necessitated
investigation of peculiarities that technologically important processes
may exhibit at this scale.\cite{packan_2000,ho_controlled_2008} One such
process that plays a key role in the doping of semiconductor chips is
the diffusion of impurities in elemental crystalline hosts.\cite{rmp}
In experiments of Cowern et al.\cite{cowern1990,cowern1991} it was
established that some dopants in the silicon host exhibit exponential
diffusion profiles instead of the conventional Gaussian ones. This
seemingly non-Fickian behavior was attributed to the diffusion mediated
by a mobile intermediate state of impurities induced by interaction with
point defects, such as vacancies and interstitials.  A phenomenological
model based on the notion of the mobile state developed by Cowern et
al.\cite{cowern1990,cowern1991} satisfactorily described the experimental
observations with the use of only two adjustable parameters.  However,
the development of the model into a quantitative theory
has been hampered by insufficient understanding of the microscopic
mechanisms underlying the mobile state.  The main problem poses the
presence in semiconductor hosts of several competing mechanisms with
their relative importance being impurity-specific and not fully
understood.\cite{rmp,PhysRevLett.62.1049,cowern1990,cowern1991,%
packan_2000,jung2004pair,general_expression2010,step_profile,%
v-Icontrovercy_eq_vs_neq,mirabella_mechanisms_2013}

The analysis of defect-assisted diffusion considerably simplifies when
only one mechanism dominates, as was the case in the diffusion in the
Cu(001) surface layer where exponential tails in diffusion profiles were
also observed.\cite{In-V-attraction,In/cu(001)} The vacancy mechanism was
shown to be dominant\cite{In-V-attraction,In/cu(001),Co/cu(001),Pd/Cu}
and due to its simplicity it was possible not only to reliably fit
the values of microscopic parameters to experimental data but also to
confirm them in the first-principles calculations.\cite{Pd/Cu} Diffusion
in two dimensions (2D), however, is qualitatively different from 3D
case\cite{Brummelhuis1988,toroczkai1997,Co/cu(001)} and so cannot serve
as a model of vacancy-mediated diffusion in 3D systems.

But the vacancy mechanism is also common in 3D solids, even
in semiconductors.\cite{Ge_E_b_ab_ini,mirabella_mechanisms_2013}
Arguably, it has been best studied in the FCC metals in the framework of
the classic five frequency model (5FM).\cite{lidiard,LECLAIRE1978}
There exists a wealth of literature on the pertinent host-impurity
systems, on approximate solutions of the model, on comparison with
experiment as well as large databases containing systematized data
on both first-principles calculations of the model parameters
and on their empirical values (see, e.\ g., Refs.\
\onlinecite{manning,philibert,voglAlFe,PhysRevB.43.9487,%
LDA+UinAl,Wolverton2007,ab_init_Al2009,entropies_in_Al,%
point_defects_ab_initio2014,wu_high-throughput_2016,Wu2017,%
vacanciesDB2014} and references therein).

Furthermore, the mobile state of impurity in the 5FM is also
known.\cite{lidiard,krivoglaz1961,krivoglaz-repetskiy,voglAlFe} It
appears in the case of strong impurity-vacancy (I-v) attraction when
the I-v pairs that form can exercise long sequences of diffusion
jumps because in the bound state the diffusion-mediating vacancy
is permanently available.  Similarly to the 2D case discussed
above,\cite{In-V-attraction,In/cu(001),oshanin} this mechanism was
shown to produce non-Gaussian diffusion profiles (NGDPs) also in
the FCC systems.\cite{cond-mat/0505019} The term NGDPs will be used
throughout the paper to refer to the profiles that Cowern et al. called
exponential\cite{cowern1990,cowern1991} because unlike in the surface
layer where truly exponential behavior can be observed due to the
specifics of the STM experiment,\cite{In-V-attraction,In/cu(001)} below
we will see that in 3D the tails of the profiles are asymptotically
always Gaussian with the exponential behavior being observable only at
intermediate distances.

The aim of the present paper is to develop a quantitative theory of
impurity diffusion propagated by the bound I-v pairs using the framework
of the 5FM for FCC hosts. The theory will be essentially based on the
phenomenological approaches of Refs.\ \onlinecite{cowern1990,In/cu(001)}
with the input parameters calculated within the 5FM from the data
available in literature sources.\cite{manning,PhysRevB.43.9487,%
LDA+UinAl,Wolverton2007,ab_init_Al2009,entropies_in_Al,%
point_defects_ab_initio2014,wu_high-throughput_2016,Wu2017}

The paper is organized as follows. In the next section the 5FM
is briefly introduced; in Sec.\ \ref{Iv} theory of diffusion
of individual tightly-bound I-v pairs is presented; in Sec.\
\ref{phenomenological_theory} the diffusion of the ensemble of immobile
impurities and mobile pairs is treated within a phenomenological
theory based on the approach of Cowern et al.\cite{cowern1990}; in Sec.\
\ref{d-profiles} the integro-differential equation describing 3D diffusion
of impurities mediated by the mobile state is derived and illustrated
by simulation of 1D NGDPs for lanthanum impurity in aluminum host;
in the concluding section \ref{conclusion} the results obtained are
briefly summarized.
\section{\label{the-model}The 5FM}
\begin{figure}
\begin{center}
\includegraphics[viewport = 0 20 229 206, scale = 0.45]{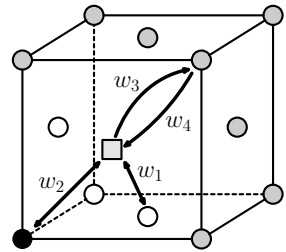}
\end{center}
\caption{\label{5fm}Jump frequencies of the vacancy (gray square)
in the vicinity of the impurity (black circle) as defined in the
5FM: $w_1$--the jumps in the first coordination shell (CS) of the
impurity; $w_2$--frequency of exchange with the impurity atom;
$w_3$--frequency of dissociative jumps away from the impurity into
higher CS; $w_4$--associative jumps from higher CS into the first one.
Not shown in the figure is the frequency $w_0$ of exchange with atoms
in the host bulk.} \end{figure}

The 5FM for the FCC
lattice\cite{lidiard,manning1962,LECLAIRE1978,manning,philibert} is a
representative of the class of models that describe the vacancy-mediated
diffusion as a stochastic process characterized by a set of
the transition rates or frequencies of the vacancy jumps
between the sites of the host lattice. In Fig.\ \ref{5fm} the meaning of
the five frequencies $w_k$, $k=0-4$, for the FCC host are explained. In
the canonical model that will be used in the present paper the
vacancy jumps are restricted to only nearest neighbor (NN) sites though
generalizations on more complex models with larger sets of parameters are
possible.%
\cite{LECLAIRE1978,manning,NNN_jumps,Bocquet,wu_high-throughput_2016,nist} 
As can be seen from the definition of 5FM in Fig.\ \ref{5fm},
in the vicinity of impurity only one vacancy is assumed to be present.
This, of course, is an approximation but it will be sufficient for our
purposes because the NGDPs we are going to study are
the most pronounced at low temperatures where the vacancy concentration
is very small.  For example, using the experimental vacancy formation
enthalpy in aluminum\cite{wu_high-throughput_2016}
$E_f=0.67$,
the vacancy concentration at temperature $50\,^\circ\mbox{C}$ that will be
used in our simulations in Sec.\ \ref{d-profiles} can be estimated to be
\begin{equation}
	c_v\simeq e^{-E_f/k_BT}=3.6\cdot10^{-11}.
	\label{c_v}
\end{equation}
Here we neglected the entropic contribution (in a rigorous
treatment the Gibbs free energy should be used instead of the
energy) because the formation entropies $s_f$ are usually quite small
$s_f\approx1$\cite{vacancies_S} and the temperature $50\,^\circ\mbox{C}$
in energy units is $\sim0.03$~eV while the errors in both experimental and
theoretical definitions of $E_f$ are considerably larger being at least of
order of 
0.1~eV.\cite{vacancies_S,vacanciesDB2014,ab_init_Al2009,wu_high-throughput_2016}

The jump frequencies can be calculated from the values of the
activation barriers $E_k$ and the attempt frequencies $\nu_k$ (see Table
\ref{table1}) as\cite{Wu2017}
\begin{equation}
	w_k=\nu_ke^{-E_k/k_BT}.
	\label{w_k}
\end{equation}
Because the stochastic dynamics in the 5FM is governed by thermal
excitations, it satisfies the detailed balance condition which establishes
the following relation between the frequencies and the binding energy
$E_b$ of the I-v pair
\begin{equation}
	\frac{w_3}{w_4}\simeq\exp\biggl(-\frac{E_b}{k_BT}\biggr).
	\label{w3/w4}
\end{equation}
In this definition $E_b$ is assumed to be positive for I-v
attraction.  As in Eq.\ (\ref{c_v}), here we also neglected the entropy
contribution because of its smallness in comparison with the errors in
$E_b$.\cite{entropies_in_Al,wu_high-throughput_2016}
\begin{table}
\caption{\label{table1}Parameters entering Eqs.\ (\ref{w_k}) and
(\ref{DIH}) corresponding to lanthanum impurity (I) in aluminum host 
(H) taken from the database of Ref.\ 
\onlinecite{Wu2017}}
	\begin{tabular}{c|c|c}
		\hline
		$k$&$E_k$(eV)&$\nu_k$(THz)\\
		\hline
0&0.5814& 4.2242\\
1&1.18& 3.9868\\
2&0.0623& 2.5182\\
3&0.5637& 4.1975\\
4&0.092& 4.0664\\
\hline
&$Q$(eV)&$D_{0}$(cm$^2$/s)\\
\hline
I&0.7776&0.081112\\
H&1.2661&0.064623
	\end{tabular}
\end{table}

The frequencies $w_0$--$w_4$ together with the vacancy concentration
from Eq.\ (\ref{c_v}) and the value of the lattice parameter $a$  can
be used to calculate the diffusion constant of the impurity in the 5FM
as\cite{lidiard,manning1962,LECLAIRE1978,manning,philibert}
\begin{equation}
	D=c_vw_2\frac{w_4}{w_3}fa^2,
	\label{D_M}
\end{equation}
where the correlation factor 
\begin{equation}
f =\frac{2w_1+7F_3(w_4/w_0)w_3}{2w_1+2w_2+7F_3(w_4/w_0)w_3}.
	\label{f_manning}
\end{equation}
Accurate expressions for $F_3$ were derived in Refs.\
\onlinecite{manning1964,koiwa_5fm_f}.  For our purposes will be
necessary the values of $f$ in the case of strong I-v binding
(large $E_b$). According to Eq.\ (\ref{w3/w4}) this corresponds
to cases when $w_3\to0$ and/or $w_4\to\infty$. From Eq.\
(\ref{f_manning}) it is seen that in the first case ($w_3\to0$) the
value of $F_3$ is irrelevant while in the second case it is known
exactly:\cite{lidiard,manning1962,LECLAIRE1978,manning,philibert}
\begin{equation}
	F_3(w_4/w_0\to\infty)=2/7.
	\label{F_3infty}
\end{equation}
Thus, the correlation factor in the limit of infinitely strong binding is
\begin{equation}
	f_\infty = \frac{w_1+w_3}{w_1+w_2+w_3}.
	\label{f_infty}
\end{equation}
Eq.\ (\ref{f_manning}) also accurately reproduces the exactly
known value for the case of self-diffusion when all $w_i$ are equal:
\begin{equation}
	f_0\simeq0.781.
	\label{f0}
\end{equation}

The most serious obstacle to quantitative predictions based on the 5FM
is that the parameters of the model either as fitted to experimental
data or as obtained in first-principles calculations may contain quite
significant errors.\cite{DU2003140,ab_init_Al2009,Wu2017,vacanciesDB2014,%
wu_high-throughput_2016} The problem aggravates at lower temperatures
because of the Arrhenius behavior in Eq.\ (\ref{w_k}).  In order to
reconcile the measured and calculated values of th host ($H$) and the impurity
($I$, though this superscript will be usually omitted for brevity) diffusion 
constants 
\begin{equation}
	D^{H,I}=D_0^{H,I}\exp\left(-\frac{Q^{H,I}}{k_BT}\right),
	\label{DIH}
\end{equation}
a correction coefficient was suggested in Ref.\
\onlinecite{wu_high-throughput_2016}
\begin{equation}
	C_{shift}=A_{shift}\exp\left(-\frac{E_{shift}}{k_BT}\right).
	\label{Cshift}
\end{equation}
In the case of aluminum $A_{shift}=12$ and $E_{shift}=0.2$~eV, so
at $50\,^\circ\mbox{C}$ the coefficient $C_{shift}\approx10^{-2}$
which means that the discrepancy between the theory and the experiment
amounts to two orders of magnitude at this temperature.  Because various
quantities in the 5FM may depend on six parameters (five frequencies
plus the vacancy formation energy) and their errors combine, we will
try to maximally reduce their number in our calculations and whenever
possible use instead experimentally measurable quantities.
\section{\label{Iv}Diffusion of tightly bound I-v pairs} 
According to Eq.\ (\ref{w3/w4}), strong I-v binding meaning large
$E_b$ value takes place when either $w_3$ is very small or $w_4$ is very large
or both.  The case of vanishing escape frequency $w_3$ was solved in
Ref.\ \onlinecite{lidiard} so we will consider a more general case
of large $w_4\gg w_3$  which comprises also the small $w_3$
case.  

\begin{figure}
\begin{center}
\includegraphics[viewport = 0 30 400 400, scale = 0.4]{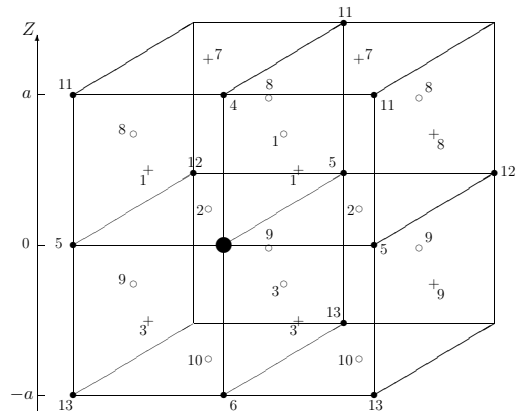}
\end{center}
\caption{\label{geometry}Axisymmetric diffusion along $Z$ direction of the 
FCC lattice as described by the 5FM: Distribution of lattice sites in 
the vicinity of the impurity (black circle) among 13 equivalence classes.
Black points---the classified sites in the cubes vertexes; crosses---the
face centered sites on the visible faces in the drawing, open circles---on
the invisible faces.}
\end{figure}

To begin with, let us find the solution in the limit
of the infinitely large of $w_4$ and for
definiteness let us restrict our consideration only to diffusion along $Z$
direction because in cubic lattices all $\langle100\rangle$ directions
are equivalent.  Similar to Ref.\ \onlinecite{lidiard},
the problem in this case can be reduced to the solution of a set of
three equations for the probabilities of three inequivalent mutual I-v 
orientations.  They correspond to the vacancy being in one of three 
classes 1--3 of the sites in the first coordination shell (CS) of the 
impurity (see Fig.\ \ref{geometry}); inside these classes all positions
are equivalent due to our choice of the symmetry direction. In Fig.\ 
\ref{geometry} all sites pertinent to the 5FM
are divided into 13 equivalence classes. In the case of infinitely large
$w_4$ the vacancy will spend all its time on the sites in classes 1--3
because of the following.  Classes from 4 to 13 comprise the sites that can
be reached from the first CS in one jump with the rate $w_3$.  If the vacancy
jumps at one of these sites, it will have at least one NN site in the
first CS (the one it just jumped from) but in general there will be $l\ge1$
such sites. At the next step the vacancy either returns back to the
first CS with the rate $lw_4$ or diffuses further away from the impurity 
with the rate $(12-l)w_0$. The probability of jumping back to the first CS 
is given by the ratio\cite{n-fold}
\begin{equation}
	p_{back}=\frac{lw_4}{lw_4+(12-l)w_0}
	\label{p_back}
\end{equation}
which tends to unity when $w_4\to\infty$.  Moreover, because the
residence time $\Delta t$ of the vacancy at the ``outer'' site is
inversely proportional to the total rate in the denominator of Eq.\
(\ref{p_back}) $\Delta t\propto 1/[lw_4+(12-l)w_0]$, as $w_4\to\infty$
the time goes to zero. Thus, the vacancy spends all its time in the
first CS of impurity with the excursions to sites in classes 4-13 
serving only to jumps between classes 1-3 in the first CS.

To formalize this picture let us number the (001) planes in the
direction of the diffusion by integer numbers $m,n$ ranging from minus
to plus infinity. When the impurity is positioned in plane $n$
the sites of class 1 are in the plane $m=n+1$, sites of class 2 are in
the same plane ($n$) and those of class 3 in the plane $m=n-1$ (see
Fig.\ \ref{geometry}).  The master equation describing the evolution
of the I-v pair can be written with the use of the transition matrix
$\tilde{W}^{\alpha\alpha^\prime}_{nm}$ as
\begin{equation}
\frac{d}{dt}C^{\alpha}_{n}(t)=\sum_{m,\alpha^\prime}
\tilde{W}^{\alpha\alpha^\prime}_{nm}C^{\alpha^\prime}_{m}(t),
\label{I-v}
\end{equation}
where $C^{\alpha}_{n}$ is the concentration of pairs on plane $n$ with 
the vacancy being in class $\alpha=1-3$.
The transition matrix that has the standard loss-gain 
structure\cite{van_kampen} is
\begin{widetext}
\begin{equation}
\tilde{W}_{nm}=\begin{bmatrix}
	-(2w_1+w_2+\frac{7}{4}w_3)\delta_{nm} & (2w_1+\frac{3}{2}w_3)\delta_{nm}&w_2\delta_{m-n,1} +\frac{1}{4}w_3\delta_{nm}\\
(2w_1+\frac{3}{2}w_3)\delta_{nm}&-(4w_1+3w_3)\delta_{nm}&2(w_1+\frac{3}{2}w_3)\delta_{nm}\\
w_2\delta_{n-m,1}+\frac{1}{4}w_3\delta_{nm}&(2w_1+\frac{3}{2}w_3)\delta_{nm}&-(2w_1+w_2+\frac{7}{4}w_3)\delta_{nm}
\end{bmatrix},
\label{real_space}
\end{equation}
\end{widetext}
where the rows and columns are numbered by the class indexes
$\alpha,\alpha^\prime$ while subscripts $m$ and $n$ denote the planes.
The non-diagonal entries of the matrix correspond to transition rates
between the classes.  For example, $\tilde{W}_{nm}^{21}$ is the rate
of the vacancy jumps from a site in class 1 to a site in class 2 as
well as a possible ensuing displacement of the impurity from plane $m$
to plane $n$. The jump can proceed by several routes.  The vacancy can
reach class 2 directly via NN jumps. This contributes $2w_1$ to the
rate because there are exactly two sites in class 2 that are NN to a
site from class 1. The indirect way is to first jump out of the first
CS on one of two NN sites in class 8 or on the site in class 5 (in total
three possibilities). But the return jumps will end up in class 2 only in
half of the cases which reduces the contribution to $3w_3/2$. The matrix
element is proportional to $\delta_{mn}$ because the vacancy moves do
not change the impurity position. In fact, only direct I-v exchanges
lead to the impurity diffusion, so the only contributions that change
the position of the impurity are those proportional to $w_2$, which is
reflected in Eq.\ (\ref{real_space}) (see Fig.\ \ref{5fm}).

To simplify the task of solving Eq.\ (\ref{I-v}), let us restrict our
attention only to macroscopic diffusion that develops at a spatial
scale much larger than the lattice constant $a$. At this scale
the positions of the impurity and the nearby vacancy are essentially the
same so concentrations $C_n^{\alpha}$ need not be distinguished and only
evolution of the total pair density
\begin{equation}
C_n=\sum_{\alpha=1}^3C_n^\alpha  \label{Cn} 
\end{equation} 
will be detectable experimentally via the impurity profiles.  Long
diffusion distance means long duration.  But solutions of stochastic
evolution equations of the kind of Eq.\ (\ref{I-v}) tend toward the
equilibrium via exponentially attenuating modes $\propto\exp(z_it)$
with all $z^p_i$ being less or equal to zero. So we will be interested
only in the longest-lived modes that correspond to the smallest $z_i$.

For a translationally-invariant system with constant coefficients the
standard way to find the attenuation rates is to reduce differential
equation Eq.\ (\ref{I-v}) to an algebraic equation with the use of
integral transforms which we chose to be the Laplace transform in time
variable and the Fourier transform in the spatial variables (the LF
transform). The 1D Fourier transform of the transition matrix Eq.\
(\ref{real_space}) is
\begin{equation}
	\tilde{W}_{K}=\begin{bmatrix}-x-y & 
	x&y+\gamma_K \\
	x&-2x&x\\
	y+\gamma_K^*&x&-x-y 
\end{bmatrix},
\label{k-space}
\end{equation}
where
\begin{eqnarray}
	\label{xygam}
	x&=&2w_1+\frac{3}{2}w_3\\
	y&=&w_2+\frac{1}{4}w_3\\
	\gamma_K&=&w_2(e^{iaK/2}-1)
\end{eqnarray}
and $a/2$ is the distance between successive (001) planes.
Taking further the Laplace transform over $t$ with the parameter $z$ 
we obtain the characteristic equation 
\begin{equation}
	\det(z-\tilde{W}_{K})=0.
	\label{characteristic1}
\end{equation}
As explained above, we do not need exact expressions for the solutions
of this equation but are interested only in the smallest eigenvalue that
describes the pair diffusion. At small Fourier momenta $K$ corresponding
to large distances in real space the eigenvalue will tend to zero as
$\sim K^2$ because the matrix in Eq.\ (\ref{k-space}) is Hermitian.
The smallest root (let us denote it $z_1$) can be easily found to this
accuracy from Eq.\ (\ref{characteristic1}) from the linearized equation
because as $z_1\to0$ the higher orders in $z$ can be dropped:
\begin{equation}
z_1^{p}({\bf K}) \simeq \frac{w_2(w_1+w_3)}{12(w_1+w_2+w_3)}a^2{\bf K}^2.
	\label{z1}
\end{equation}
Here superscript $p$ reminds us that the solution describes the pair
diffusion and the momentum $(0,0,K)$ describing the diffusion along $Z$
direction is replaced by a general momentum ${\bf K}=(K_X,K_Y,K_Z)$
because in cubic crystals the diffusion is isotropic to this order in $K$.
The remaining two eigenvalues are finite at small $K$ and up to terms
$O(K^2)$ are
\begin{eqnarray}
	\label{z23}
	z_2^{p}({\bf K}) &=&-3x+O(K^2)\simeq-6w_1-4.5w_3\nonumber\\	
	z_3^{p}({\bf K}) &=&-x-2y+O(K^2)\simeq -2(w_1+w_2+w_3)
\end{eqnarray}
as can be easily verified by direct substitution in Eq.\
(\ref{characteristic1}) at $K=0$. As is seen, the eigenvalues 2--3
remain finite as $K\to0$. Therefore, only the term corresponding to
$z_1^{p}({\bf K})=O(K^2)$ will survive at large times so the Fourier
transform of the pair density Eq.\ (\ref{Cn}) will behave as
\begin{equation}
C_{\bf K}(t)|_{t\to\infty}\propto 
\exp{\left[-z_1^{p}({\bf K})t\right]}.
	\label{Ct2infty}
\end{equation}
Differentiating this by $t$ and taking the inverse Fourier transform
with respect to the spatial variables bring about the conventional
diffusion equation
\begin{equation}
\partial C({\bf R},t)/\partial t\simeq D_m\nabla^2 C({\bf R},t),
	\label{DEI-v}
\end{equation}
where $C$ is the continuum approximation to $C_n$ and according to Eqs.\
(\ref{z1}), (\ref{Ct2infty}), and (\ref{f_infty}) the diffusion constant
of the I-v pairs is
\begin{equation}
	D_m=\frac{w_2(w_1+w_3)a^2}{12(w_1+w_2+w_3)}
=\frac{w_2}{12}f_\infty a^2.
	\label{Dm}
\end{equation}
For $w_3=0$ this expression coincides with the result of Ref.\
\onlinecite{lidiard}.  

The most important property of $D_m$ is that in contrast to the impurity
diffusion constant Eq.\ (\ref{D_M}), it is not proportional to the
vacancy concentration and so can be much larger than $D$.  Assuming strong
binding in which case one can use $f_\infty$ as $f$ in Eq.\ (\ref{D_M})
with the use of Eqs.\ (\ref{Dm}), (\ref{c_v}), and (\ref{w3/w4}) one gets
\begin{equation}
	\frac{D_m}{D}\simeq \frac{w_3}{12c_vw_4}
	\simeq\frac{1}{12}e^{(E_f-E_b)/k_BT}.
	\label{Dm/D}
\end{equation}
Simple bond-counting arguments suggest that the vacancy formation energy
$E_f$ should be larger than the binding energy $E_b$.  The latter can be
found by comparing the energies of a vacancy surrounded by only host atoms
and the vacancy with one neighbor replaced by the impurity which amounts
to the difference between energies of a single atomic bond. Creation
of a vacancy, on the other hand, costs about six atomic bonds in the
FCC lattice.  So as $T\to0$ the ratio Eq.\ (\ref{Dm/D}) normally can 
take arbitrarily
large values.  For example, in the case of the La\underline{Al}
system that will be used in illustrative calculations below, the ratio in
Eq.\ (\ref{Dm/D}) at temperature $50\,^\circ\mbox{C}$ may reach according
to Eqs.\ (\ref{c_v})--(\ref{w3/w4}) and Table \ref{table1} the value
\begin{equation}
	({D_m}/{D})_{T=50\,^\circ\mbox{\scriptsize C}}\approx 105.
	\label{Dm/D50C}
\end{equation}
It is the presence of two modes of impurity diffusion with very
different diffusion constants that underlies the phenomenon of NGDPs.
\subsection{\label{decay-rate}Decay rate of the I-v pair} 
Despite large diffusivity, the bound I-v pairs cannot diffuse too
far from the place of their association because of their finite lifetime.
Irrespective of how strong the I-v binding is, from Eq.\ (\ref{w_k})
it is seen that at finite temperature neither $w_3$ can be strictly
equal to zero nor $w_4$ can be infinitely large.  Thus, if the lifetime
of the pair is equal to $\tau_{decay}$ its decay rate $r=1/\tau_{decay}$
and the characteristic distance $\lambda$ of the pair diffusion before
its decay is equal (up to a numerical constant) to the diffusion length
of the pair during its lifetime
\begin{equation}
	\lambda = \sqrt{D_m\tau_{decay}}=\sqrt{D_m/r}.
	\label{lambda}
\end{equation}
In Refs.\ \onlinecite{cowern1990,cowern1991} this quantity is called
the mean projected path length.

In the 5FM the decay rate $r$ can be found as the rate of definite
separation of the vacancy from the impurity. It is often approximated by
the rate of escape from the first CS into higher coordination shells which
is equal to $7w_3$\cite{krivoglaz-repetskiy,voglAlFe} because there is
seven NN sites to a site in the first CS that do not belong to the first
CS, as can be seen from Figs.\ \ref{5fm} or \ref{geometry}. However, as
we already saw, this approximation can be completely misleading in the
case of large $w_4/w_0$ ratio when the vacancy returns into the first
CS with probability that almost equals to unity.  To account for this
$r$ can be represented as a product of the ``bare'' escape rate $7w_3$
and the renormalization factor $p_{\infty}$ equal to the probability of
definite I-v dissociation when the vacancy diffuses infinitely far away
from the impurity
\begin{equation}
	r= 7w_3p_\infty.
	\label{r}
\end{equation}

\begin{figure}
\begin{center} 
	\includegraphics[viewport = 100 20 229 206, scale = 0.75]{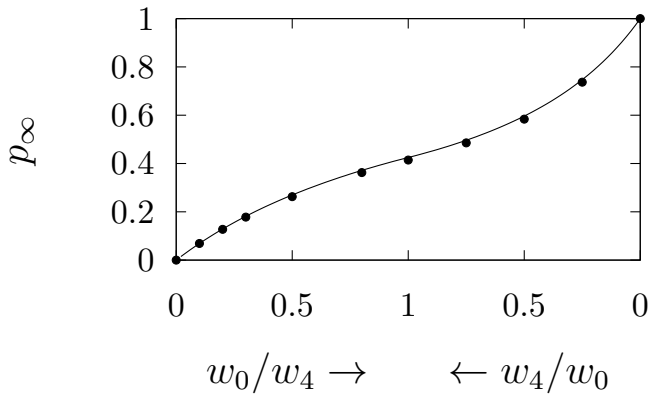} 
\end{center} 
\caption{\label{runaway_p}Probability for the vacancy to definitely leave
the impurity neighborhood after escape from the first CS:
bullets---the KMC simulation data and the exactly known end points;
solid line: Eq.\ (\ref{p_approx}).}
\end{figure}

The problem of calculating the vacancy escape probability $p_\infty$
is equivalent to finding the return probability into the first CS which
is a standard problem of the random walk theory (see, e.\ g.,
Ref. \onlinecite{diff_from_sphere} and references therein).  But to
avoid complicated combinatorial calculations we will assess $p_\infty$
with the help of numerical simulations.  The escape of the vacancy
from the first CS into the space beyond sufficiently large radius 
$R_{max}$ was simulated for several values of the ratio $w_4/w_0$ and for
two radii $R_{max}=100(a/2)$ and $300(a/2)$ 
and then interpolated to $R_{max}=\infty$. The results are
presented in Fig.\ \ref{runaway_p} together with the exactly known two
end point values $p_\infty(w_4/w_0=0)=1$ and $p_\infty(w_4/w_0=\infty)=0$
and with the approximating interpolating expression
\begin{equation}
	p_\infty\simeq (1+bw_4/w_0)^{-1},
	\label{p_approx}
\end{equation}
where $b\approx1.35$ was found from the largest simulated ratio
$w_4/w_0=10$ because large $w_4/w_0$ values are the most interesting to
our purposes. However, because of the diminishing number of the decays
as $w_4/w_0\to\infty$, the statistics are difficult to gather when the
ratio is in the range $O(10^4-10^7)$ that we are interested in (see the
next paragraph).  It can be calculated in this case within the rigorous
approach of Ref.\ \onlinecite{cond-mat/0505019}, as will be shown in
the subsequent publication.\cite{2follow} Therefore, in our calculations
below we will use this more accurate value $b=1.32$.

Now substituting Eqs.\ (\ref{r}) and (\ref{p_approx}) into Eq.\
(\ref{lambda}) we can obtain an explicit expression for the
phenomenological parameter $\lambda$ of the theory of Ref.\
\onlinecite{cowern1990} in terms of the frequencies of the 5FM.
To calculate $\lambda$ for 50 impurities listed in the database of
Ref.\ \onlinecite{Wu2017} for aluminum host use has been made of
Eqs.\ (\ref{lambda}), (\ref{Dm}), (\ref{r}) and (\ref{p_approx}).
Nine perspective impurities with the largest values of $\lambda$ were
identified (see Table \ref{table2}).  It turned out that in all nine
cases the ratio $w_4/w_0$ were very large, of order $O(10^4-10^7)$.
\begin{table}
\caption{\label{table2}Impurities in aluminum host with $\lambda>10a$
at temperature $T=50\,^\circ\mbox{C}$ calculated from Eqs.\
(\ref{lambda2infty}) and (\ref{DIH}) with the parameters taken from Ref.\
\onlinecite{Wu2017}; $E_b=E_3-E_4$.}
\begin{tabular}{l|r|c}
\hline
I& $\lambda$ (nm)&$E_b$ (eV)\\
\hline
La & 328 &  0.47\\
Se & 161 &  0.44\\
Te & 34 &   0.46\\
Bi & 30 &   0.41\\
Nd & 22 &   0.27\\
Ce & 21 &   0.26\\
Pb & 17 &   0.36\\
Tl & 8.2&   0.31\\
Sb & 4.5&   0.30
\end{tabular}
\end{table}

The impurities with large $\lambda$ are the most interesting ones from
experimental perspective, so it is desirable to calculate the parameter
with maximum precision. But, as was discussed at the end of the previous
section, the accuracy of the 5FM frequencies values is currently rather
modest and the errors may compound in the uncertainty of the value of
$\lambda$. So it would be preferable to express the latter in terms
of more reliable parameters.  In the case $w_4\to\infty$ this can be
achieved as follows.  At large $w_4$
\begin{equation}
	p_\infty|_{w_4\gg w_0}\simeq\frac{w_0}{bw_4}.
	\label{p,w42infty}
\end{equation}
Substituting this into Eq.\ (\ref{r}) and using the decay rate thus obtained 
in the ratio $D_m/r$, with the use of Eq.\ (\ref{Dm}) one gets after some 
rearrangement
\begin{equation}
\frac{D_m}{r}=\left(\frac{bf_0}{84}\right)
\frac{c_vw_2({w_4}/{w_3})f_\infty a^2}{c_vw_0f_0a^2}a^2.
	\label{Dm/r}
\end{equation}
Comparing this with Eq.\ (\ref{D_M}) and using Eq.\ (\ref{f0}) from Eq.\ 
(\ref{lambda}) one arrives at the expression
\begin{equation}
	\lambda|_{w_4\gg w_0}=Aa\sqrt{\frac{D}{D^H}},
	\label{lambda2infty}
\end{equation}
where 
\begin{equation}
	A=\sqrt{bf_0/84}\approx0.111
	\label{A}
\end{equation}
is a numerical constant. Thus, the parameter $\lambda$ of the phenomenological theory in the large-$w_4$ case can be calculated from only experimentally 
measurable quantities.
\subsection{\label{unstable-diff}Diffusion profiles of unstable I-v pairs}
One consequence of the pairs instability is that their diffusion cannot
be described by the conventional diffusion equation Eq.\ (\ref{DEI-v})
because the second Fick's law expresses the conservation of the diffusing
particles which is not the case with the pair diffusion. Being unstable,
the I-v pairs obey instead of Eq.\ (\ref{DEI-v}) the non-Fickian diffusion 
equation suggested in Ref.\ \onlinecite{In/cu(001)}:
\begin{equation}
\partial G_p({\bf R},t)/\partial t= D_m\nabla^2 G_p({\bf R},t)-rG_p({\bf R},t),
	\label{gastel_eq}
\end{equation}
where we introduced the pair Green's function (GF) that satisfies the
delta-function initial condition
\begin{equation}
	G_p({\bf R},t=0)=\delta({\bf R})
	\label{G(t=0)}
\end{equation} 
and describes the probability to find the pair at point ${\bf R}$ at
time $t$.  Integrating Eq.\ (\ref{gastel_eq}) over the space variables it
is seen that the probability to find the pair at time $t$ anywhere in the
system diminishes as $e^{-rt}$, as expected. 
Explicit expressions for the GF of the decaying pair in the space-time
variables can be written straightforwardly for any dimension $d$ 
\begin{equation} 
	G_p({\bf R},t)=\frac{1}{(4\pi D_mt)^{d/2}}
	\exp\left(-\frac{{\bf R}^2}{4D_mt}-rt\right).
\label{G_m}
\end{equation}
Below we will also need the LF-transformed $G_p$ that is also easily
found from Eqs.\ (\ref{gastel_eq}) and (\ref{G(t=0)}) as
\begin{equation}
	G_p({\bf K},z)=\frac{1}{z+r+D_m{\bf K}^2}.
	\label{GpKz}
\end{equation}
Despite instability of the pairs, the impurity density should conserve
irrespective of the diffusion mechanism.  This is indeed the case
if we take into account the impurities from the decayed pairs that
simply immobilize (become stable) and their density grows with time
as\cite{In/cu(001)}
\begin{equation}
	G_{sp}({\bf R},t) = r\int_0^t dt^\prime  G_p({\bf R},t^\prime).
	\label{G_sp}
\end{equation}
Integrating this over the spacial variables is easy to check
that the normalization of the sum 
\begin{equation}
	G_p({\bf R},t)+G_{sp}({\bf R},t)
	\label{Gp+Gsp}
\end{equation}
is equal to unity at all $t$. 

We note that the first term in Eq.\ (\ref{Gp+Gsp}) at large time goes
to zero so the impurity distribution is dominated by the second term.
Due to the specifics of the STM technique, only the second term was
observed experimentally in Refs.\ \onlinecite{In-V-attraction,In/cu(001)}
and only at $t=\infty$ in which case the first term vanished and the
second acquired the exponential asymptotic behavior in the spatial
variables\cite{In-V-attraction,In/cu(001)}
\begin{equation}
	G_{sp}|_{|{\bf R}|\to\infty}\propto\exp\big(-|{\bf R}|/\lambda\big). 
	\label{exponential}
\end{equation}
It can be shown that this is a universal behavior at $t=\infty$ in all 
dimensions. To see this we first notice that the Laplace transform of 
a function at $z=0$ is just the integral of the function over the time 
variable from zero to infinity.  Thus, using Eqs.\ (\ref{GpKz}) and
(\ref{G_sp}) one gets with the use of the inverse Fourier transform
\begin{eqnarray}
	G_{sp}({\bf R},\infty)&=&\frac{r}{(2\pi)^d}\int 
	d^d{\bf K}\frac{e^{i{\bf K\cdot R}}}{r+D_m{ \bf K }^2}\nonumber\\
&=& \frac{1}{(2\pi)^d}\int d^d{\bf K}\frac{e^{i{\bf K\cdot R}}}
{1+(\lambda{ \bf K })^2} \equiv G^P({\bf R}),
	\label{poisson}
\end{eqnarray}
where on the second line we introduced the kernel of the screened Poisson
equation\cite{poisson_eq} $G^P$ which is known to have the exponential
asymptotic behavior Eq.\ (\ref{exponential}) in all dimensions.

In contrast to the diffusion in surface layers where it is
possible to observe individual I-v pairs and ignore the rest of the
impurities,\cite{In-V-attraction,In/cu(001)} in 3D diffusion all impurity
atoms contribute so at $t=\infty$ the profile will span the whole crystal
and will be seen as just the homogeneous equilibrium distribution.
Therefore, NGDPs can be observed only at finite $t<\infty$ in which
case their asymptotic will be Gaussian as can be easily illustrated
in 1D geometry.  Setting in Eqs.\ (\ref{G_m}) and (\ref{G_sp}) $d=1$
and ${\bf R}=X$ and taking the integral over $t$ one gets
\begin{eqnarray}
	&&G_{sp}(X,t)=r\int_0^tG_p(X,t^\prime)dt^\prime
\nonumber\\
&&=-\frac{1}{4\lambda}\sum_{s=\pm1}se^{s|X|/\lambda}
\mbox{erfc}\left(\frac{|X|}{2\sqrt{D_m t}}
+s\sqrt{r t}\right).
	\label{deltaG}
\end{eqnarray}
Because at $x=-\infty$ erfc($x$)=2, the behavior of this expression at
$t=\infty$ is exponential in $|X|$, as expected. At finite $t$, however,
the behavior is Gaussian, as can be seen from the behavior of the
the erfc function at large values of its argument\cite{erfc}
\[
	\mbox{erfc}(x)|_{x\to\infty}\sim e^{-x^2}/(\sqrt{\pi}x).
\]
Thus, strictly speaking the exponential tails can never be observed in
3D concentration profiles, only some exponentially-looking transient
features, as will be shown in more general case of multiple-encounter
diffusion in Sec.\ \ref{d-profiles}. Still, the profile can be very close
to the exponential shape at low temperature and a sufficiently large density
of the associated pairs in the initial state. In this case the existing
pairs start to diffuse immediately while the immobile impurities need
first to enter into association with vacancies. At low temperature the
waiting time may be quite long because of the small vacancy concentration
so the impurity distribution due to the pair diffusion may advance to the
stage where it will only slightly differ from the infinite-time exponential
profile. 
\section{\label{phenomenological_theory}Impurity diffusion via 
multiple I-v encounters}
In the previous section we discussed diffusion of individual I-v
pairs. In particular, it was noted that Eqs.\ (\ref{G_sp}) and
(\ref{deltaG}) that describe the impurity distributions due to the
decayed pairs can be studied experimentally in the surface layers
by means of the STM microscopy which makes possible investigation of
each I-v encounter individually disregarding the diffusion of other
impurities.\cite{In-V-attraction,In/cu(001)} In conventional experiments
on diffusion in 3D bulk, however, such separation is not possible.
The impurities in the profile are indistinguishable and there is
no way to differentiate them according to their evolution history.
Therefore, theoretical description should take into account all possible
impurities: those belonging to the initial profile, associated impurities
in the mobile state, or the impurities that have already undergone one
or more I-v encounters. All these contributions should be accounted
for in a single diffusion profile with appropriate weights.  In Ref.\
\onlinecite{cowern1990} this problem was solved by first finding the
solution for zero and one encounter and then iterating the distribution
obtained as many times as necessary to describe the profile at a desired
stage of the evolution.  In the present paper we will essentially follow
this route but making it more formally refined.

Namely, we are going to use a technique of the many-body
theory usually referred to as the Dyson equation (see, e.\ g.,
Ref. \onlinecite{thouless2014quantum}). In this approach the problem of
repeated interactions of a particle is separated into an irreducible
and a reducible parts and the repeated iterations of the irreducible
part are simply summed up as a geometric series as
\begin{equation}
G_0+G_0\Sigma G_0+G_0\Sigma G_0\Sigma G_0+\dots=\frac{1}{G_0^{-1}-\Sigma}.
	\label{GS}
\end{equation}
Here the products are either convolutions in the space-time variables
or the usual algebraic products of the LF-transformed quantities; $G_0$
is the GF of a free particle and $\Sigma$ is the irreducible interaction
part that cannot be represented as two interactions separated by the
free propagation, as, e.\ g., in the third term on the left hand
side (l.h.s.). The practical observation that makes this approach
useful is that the combinatorial problem of of finding the sum of all
contributions that include the free propagation and a single or multiple
I-v encounters as represented in compact form on the r.h.s.\ of Eq.\
(\ref{GS}) can be fully recovered from only the first two terms on the
l.h.s.\ with the first term being known exactly.  This observation was
successfully applied to the problem of vacancy-mediated diffusion in
Refs.\ \onlinecite{tahir-kheli,tahir-kheli2}.  The authors effectively
derived expressions for $\Sigma$ in Eq.\ (\ref{GS}) for the cases of
the self-diffusion and in a two-frequency model and we are going to apply
this approach to the 5FM. We will call the irreducible part $\Sigma$
the diffusion kernel and express it through another quantity ${\cal D}$
that will be called the diffusivity as follows
\begin{equation}
	\Sigma({\bf K},z)=-{\cal D}({\bf K},z){\bf K}^2.
	\label{sigma}
\end{equation}
The free GF $G_0$ is easily found from the observation that without
interaction with the vacancies the impurity is immobile and remains
in its initial position: $G_0({\bf R},t)=G_0({\bf R},t=0)=\delta({\bf
R})$. With the use of the LF transform one easily finds
\begin{equation}
	G_0(z)=\int_0^{\infty}dt\,e^{-zt}\int d^d{\bf R}e^{-i{\bf K\cdot R}}
	\delta({\bf R})=\frac{1}{z}.
	\label{G0}
\end{equation}
The impurity GF with all I-v encounters being taken into account is obtained by
substituting the last two equations Eqs.\ (\ref{sigma}) and (\ref{G0})
into Eq.\ (\ref{GS}):
\begin{equation}
	G({\bf K},z)=\frac{1}{z+{\cal D}({\bf K},z){\bf K}^2}
\approx\frac{1}{z}-\frac{1}{z^2}{\cal D}({\bf K},z){\bf K}^2.
\label{Gcanonical}
\end{equation} 

The diffusion constant is found from the diffusivity as
\begin{equation}
	D={\cal D}({\bf K=0},z=0).
	\label{DD}
\end{equation}
But our main interest is in a nontrivial dependence of ${\cal D}$
on its arguments ${\bf K}$ and $z$ because if the diffusivity were
independent of ${\bf K}$ and $z$ the GF in Eq.\ (\ref{Gcanonical}) when
transformed to the time and space variables would be strictly Gaussian
without any traces of the NGDPs we are interested in.

Thus, our goal is to find explicitly the last term in Eq.\
(\ref{Gcanonical}) and with its use to recover the complete impurity GF.
To to fulfill this goal we will use Eqs.\ (3) and (4) from Ref.\
\onlinecite{cowern1990} that in the GF notation read
\begin{eqnarray}
	\label{2modes1}
	&&\partial G_m/\partial t = D_m\nabla^2 G_m -rG_m+gG_s\nonumber\\
	&&\partial (G_s+G_m)/\partial t =\partial G/\partial t 
	= D_m\nabla^2 G_m,
\end{eqnarray}
where the total impurity GF $G=G_m+G_s$ is separated into the
mobile ($G_m$) and immobile ($G_s$) parts with the subscript
``$s$'' standing for ``static'' because in contrast to Refs.\
\onlinecite{cowern1990,cowern1991} in the 5FM the impurity is always
in the substitutional position. Parameter $g$ in Eq.\ (\ref{2modes1})
is the rate of transition of the impurity from the static to the mobile
state\cite{cowern1990} which in the 5FM is the rate of I-v association.
Because in the association participates a vacancy, the rate $g$ should be
proportional to the vacancy concentration and thus is of order $O(c_v)$.
Its calculation will be discussed below.

The LF-transformed set of Eqs.\ (\ref{2modes1}) is
\begin{eqnarray}
&&(z+r+D_m{\bf K}^2)G_m({\bf K},z)-gG_s({\bf K},z)=12c_{NN}\nonumber\\
	\label{eq2}
	&&(z+D_m{\bf K}^2)G_m({\bf K},z)+zG_s({\bf K},z)=1,
\end{eqnarray}
where we assumed the initial conditions 
\begin{eqnarray}
	\label{ini}
	&&G_m({\bf R},t=0)=12c_{NN}\delta({\bf R})\nonumber\\
	&&G_s({\bf R},t=0)=(1-12c_{NN})\delta({\bf R})\nonumber\\
	&&G({\bf R},t=0)=\delta({\bf R}).
\end{eqnarray}
Here $c_{NN}$ is the vacancy concentration at the NN sites of the impurity
which can be different from $c_v$ due to the I-v interaction or because of
the way the initial state was prepared, though we will assume that it is
of the same order of magnitude as $c_v$; $12c_{NN}$ in Eqs.\ (\ref{ini})
is the density of associated I-v pairs in the initial state in the FCC
lattice which coordination number is 12 which implicitly assumes that
all sites around the vacancy are assumed to be equivalent.

It is important to note that in applying the Fourier transform we assume
that the crystal is translationally invariant which in particular means
that the rate of association $g$ is a position-independent constant. But
because $g$ depends on the vacancy concentration, this may not be
the case in experiments where non-equilibrium vacancy concentration
may acquire inhomogeneity because of their influx from the surface or
deposited layers.\cite{pantelides,v-Icontrovercy_eq_vs_neq} Such cases
cannot be treated by Eqs.\ (\ref{eq2}). It is to be understood that we
are considering the dilute systems and $c_v$ should be constant far from
the impurities in the host bulk. In the vicinity of impurity it can be
different from its bulk value due to the I-v interaction both at and out
of thermal equilibrium.  In the latter case in the Smoluchowski picture
of the vacancy capture on the impurity NN sites a non-constant vacancy
diffusion profile forms near the impurity because the NN sites serve as
the sinks.\cite{diff_from_sphere}

Because, as we pointed out in Sec.\ \ref{the-model}, the 5FM adequately
describes the I-v interaction only to first order in the vacancy
concentration, we will solve the system Eqs.\ (\ref{eq2})
only to this order by first finding the mobile GF
\begin{equation}
	G_m=\frac{12c_{NN}z+g}{z(z+r+D_m{\bf K}^2)}.
	\label{Gm}
\end{equation}
With known $G_m$ the total GF $G=G_s+G_m$ can be found 
directly from the second of Eqs.\ (\ref{eq2}) as
\begin{equation}
	G=\frac{1}{z}
	-\frac{1}{z^2}\frac{(12c_{NN}z+g)D_m{\bf K}^2}{z+r+D_m{\bf K}^2}.
	\label{G}
\end{equation}
As is seen, the approach Ref.\ \onlinecite{cowern1990} turned out to
be similar to that of Refs.\ \onlinecite{tahir-kheli,tahir-kheli2} by
giving only the zeroth and the first order terms of the expansion in Eq.\
(\ref{GS}) so to obtain the full impurity GF the Dyson equation will have
to be used.  Comparing Eq.\ (\ref{G}) with Eq.\ (\ref{Gcanonical}) we
arrive at the expression for the ${\bf K}$ and $z$-dependent diffusivity
\begin{equation}
	{\cal D}({\bf K},z)=\frac{12c_{NN}z+g}{z+r+D_m{\bf K}^2}D_m.
	\label{diffusivity}
\end{equation}
The two contribution to the second term in Eq.\ (\ref{G}) describe
somewhat different diffusion scenarios so let us discuss them separately.
The term proportional to $c_{NN}$ describes the associated I-v pairs
that are present in the initial profile. Their density $12c_{NN}$ can
be arbitrary depending on the way the profile was prepared.  From Eqs.\
(\ref{diffusivity}) and (\ref{DD}) one can see that this term does not
contribute to the diffusion constant which is natural because an arbitrary
initial condition cannot influence the quantity corresponding to thermal
equilibrium. The physical meaning of this term becomes transparent after
its rearrangement into three contributions
\begin{eqnarray}
	G|_{c_{NN}} &=& -\frac{12c_{NN}D_m{\bf K}^2}{z(z+r+D_m{\bf K}^2)}
=-\frac{12c_{NN}}{z}\nonumber\\
&+&\frac{12c_{NN}}{z+r+D_m{\bf K}^2}
+\frac{12c_{NN}r}{z(z+r+D_m{\bf K}^2)}.
	\label{DeltaG}
\end{eqnarray}
From Eq.\ (\ref{GpKz}) it can be seen that apart from the factor
$12c_{NN}$, the first term on the second line in Eq.\ (\ref{DeltaG}) is
the LF transformed $G_p$ Eq.\ (\ref{GpKz}) while the second term is Eq.\
(\ref{GpKz}) multiplied by $r/z$.  But multiplication by $1/z$ of the
Laplace transform of a function corresponds to the Laplace transform
of the integral of this function over $t$.\cite{laplace}  Thus, the
second term on the second line of Eq.\ (\ref{DeltaG}) corresponds to
LF-transformed $G_{sp}$ from Eq.\ (\ref{G_sp}), so the sum of the two
terms is Eq.\ (\ref{Gp+Gsp}) multiplied by $12c_{NN}$.  Thus, these terms
describe the diffusion profile of the pairs of density $12c_{NN}$ that
were present in the initial state. The negative term on the first line
in Eq.\ (\ref{DeltaG}) simply accounts for the fact that the associated
impurities were taken from the initial delta-function profile (see Eqs.\
(\ref{ini}) and (\ref{G0})).

In contrast to the term that describes diffusion of the I-v pairs which
already exist in the initial state, the term proportional to $g$ in Eq.\
(\ref{G}) describes the diffusion of initially immobile impurities. To
enter into the mobile state the impurities need first to be associated
with a vacancy.  This process is limited by the low vacancy concentration
and so the ensuing diffusion is much slower than the pair diffusion. The
time scales of the two diffusion modes are defined by the lifetime of
the pairs $\tau_{decay}=1/r$ and by the characteristic time of the I-v
association $\tau_{assn.}=1/g$. To compare their relative values we first
note that substitution of diffusivity Eq.\ (\ref{diffusivity}) into Eq.\
(\ref{DD}) gives
\begin{equation}
	D=\frac{g}{r}D_m
	\label{D2D}
\end{equation}
which leads to the relation
\begin{equation}
	\frac{\tau_{assn.}}{\tau_{decay}}=\frac{D_m}{D}\gg 1,
	\label{tau2tau}
\end{equation}
where the last inequality follows from Eq.\ (\ref{Dm/D}) and the
discussion that follows it. At low temperatures this ratio can be
large. In the La\underline{Al} system at $50\,^\circ\mbox{C}$ we estimated
it in Eq.\ (\ref{Dm/D50C}) as amounting to two orders of magnitude. The
main reason for this is that while the decay rate $r$ is of zeroth order
in the vacancy concentration ($O(c_v^0)=O(1)$), the association rate $g$
is of $O(c_v)$, as can be seen from the equation
\begin{equation}
	g=rD/D_m=84c_vw_4p_\infty
	\label{grD}
\end{equation}
obtained from Eqs.\ (\ref{D2D}), (\ref{Dm/D}), and (\ref{r}).

The expression for the rate of I-v association Eq.\ (\ref{grD})
was obtained in the framework of the 5FM while in Ref.\
\onlinecite{cowern1990} a Smoluchowski-type formula was used (see
their Eq.\ (12)). To compare the two approaches let us first consider
Eq.\ (\ref{grD}) in the case $w_4=w_0$ because the Smoluchowski
formula\cite{diff_from_sphere}
\begin{equation}
\Phi = 4\pi R_c \rho_0 D_v	
	\label{smoluchowski}
\end{equation}
describes the flux of vacancies of density $\rho_0$ that are being caught
by the sphere of radius $R_c$ and depends on the vacancy diffusion
constant $D_v$. The latter is defined in a continuous homogeneous
medium where all diffusion steps are equivalent which in the lattice
case means that $w_4$ cannot be different from $w_0$.  Thus, substituting
$p_\infty(w_4=w_0)=1/(1+b)$ from Eq.\ (\ref{p_approx}) in Eq.\ (\ref{grD})
we get
\begin{equation}
	g^{\mbox{\small 5FM}}(w_4=w_0)\simeq35.7 c_v w_0.
	\label{g_5FM}
\end{equation}
To find the value from the Smoluchowski formula we note that the per
volume vacancy density is $\rho_0=4c_v/a^3$ because there is four
sites in the cubic cell in the FCC lattice, the first CS
radius $R_c=a/\sqrt{2}$, and $D_v=w_0a^2$. Substituted into Eq.\
(\ref{smoluchowski}) this gives
\begin{equation}
	g^{\mbox{\small S}}\approx35.5 c_v w_0
	\label{g_Smoluchowski}
\end{equation}
in excellent agreement with the 5FM value Eq.\ (\ref{g_5FM}).

As our analysis of the database for aluminum host has revealed,
the case of large $w_4$, is particularly interesting for experimental
purposes.  To apply the Smoluchowski equation to this case we remind that
as $w_4\to\infty$ all vacancies that arrive at the sites belonging
to classes 4--13 in Fig.\ \ref{geometry} immediately form bound I-v
pair (see Sec.\ \ref{Iv}).  This means that the capture radius $R_c$
is effectively shifted toward a larger value that can be assessed
by averaging the distances from the impurity to all sites in these
classes.  Elementary calculation gives $R_c\approx1.25a$ and from Eq.\
(\ref{smoluchowski}) 
\begin{equation}
	g^{\mbox{\small S}}|_{w4\gg w_0}\approx63 c_v w_0.
	\label{g-infty-Smol}
\end{equation}
which is also very close to the 5FM value $\sim64 c_v w_0$ that can be
obtained from Eq.\ (\ref{grD}) with the use of Eq.\ (\ref{p,w42infty}).

Thus, our Eq.\ (\ref{grD}) agrees with the formula suggested in Ref.\
\onlinecite{cowern1990} in two cases where the Smoluchowski formula is
applicable but in addition covers the cases of arbitrary values of $w_4$.
But more important to us is that Eq.\ (\ref{grD}) for large $w_4$ can
be cast in the form
\begin{equation}
	g^{\mbox{\small 5FM}}|_{w4\gg w_0}=(84/bf_0)c_vw_0f_0
	=D^H/(Aa)^2,
	\label{g-infty-5fm}
\end{equation}
where $A$ is given by Eq.\ (\ref{A}). Thus, Eqs.\ (\ref{lambda2infty})
and (\ref{g-infty-5fm}) allow us to express two parameters of the
phenomenological theory only in terms of experimentally measurable
quantities.
\section{\label{d-profiles}Diffusion profiles}
In the GF approach the diffusion profiles are obtained by convolution 
of the initial profile with the impurity GF. The latter is obtained 
in our approach by first substituting the diffusivity from Eq.\ 
(\ref{diffusivity})
into Eq.\ (\ref{Gcanonical}) and then taking the inverse LF transform 
to find the GF in the space and time variables. Before proceeding with 
concrete implementation of this
procedure we have to agree on the value of the density of the I-v pairs
in the initial profile that was estimated to be equal to $12c_{NN}$.
This estimate presumes that the vacancy can be found at different NN
sites of the impurity with equal probability. That may not be the case
if the vacancies are introduced in the initial profile by means of a
non-equilibrium technique that causes non-isotropic distribution
of the vacancies on the NN positions of the 
impurities.\cite{pantelides,v-Icontrovercy_eq_vs_neq}
Such cases are beyond the scope of our approach which is restricted, as
we pointed out in Sec.\ \ref{phenomenological_theory}, to stationary and
homogeneous distributions of I-v pairs, though not necessarily corresponding
to thermal equilibrium. The latter, however, is a natural choice, so
all estimates will be done for this case. In particular, the equilibrium
density of the I-v pairs can be found from the expression
\begin{equation}
12c_{NN}^{(eq)}\simeq12c_ve^{E_b/k_BT}.
	\label{c^eq}
\end{equation}
With $E_b\approx0.47$ (see Table \ref{table1}) the associated impurities
will constitute at $50\,^\circ\mbox{C}$ about 1\% of their total
number which is a small but detectable quantity.  However, to observe the
NGDP behavior which takes place at short distances one needs to keep
initial profiles maximally sharp.\cite{cowern1990,cowern1991} But for
establishment of the equilibrium many association-dissociation events must
occur accompanied by impurity diffusion with ensuing profile smearing.
Therefore, we will assume that the initial distribution was prepared at
a temperature so low that corresponding $c_{NN}^{(eq)}$ is negligible
and that the preparation technique does not introduce excess vacancies.
So in our calculations below we for simplicity will neglect the terms
proportional to $c_{NN}$ in Eq.\ (\ref{G}). In case of necessity they
can be taken into account along the lines of derivation presented below.
Another reason for omission of these terms is that this reduces the
problem to the case studied in Ref.\ \onlinecite{cowern1990}, thus
facilitating comparison between the two approaches.

Under approximation $c_{NN}\simeq0$ the impurity GF is
\begin{equation}
	G({\bf K},z)=\left(z+\frac{gD_m{\bf K}^2}{z+r+D_m{\bf K}^2}\right)^{-1}.
	\label{GzK}
\end{equation}
This expression can be cast into the form convenient for the inverse
Laplace transform and for assessment of the relative magnitude of
different contributions:
\begin{equation}
	G({\bf K},z)=\frac{1}{z-z_1}+\frac{z_1z_2}{z_1-z_2}
	\left(\frac{1}{z-z_2}-\frac{1}{z-z_1}\right)
	\label{zz1z2}
\end{equation}
where
\begin{equation}
	z_{1,2}=-\frac{r+D_m{\bf K}^2}{2}\pm
	\sqrt{\frac{(r+D_m{\bf K}^2)^2}{4}-gD_m{\bf K}^2}
	\label{z_12}
\end{equation}
or to the leading order in $g=O(c_v)$
\begin{eqnarray}
	\label{D(K)}
	z_1&\simeq& -\frac{gD_m{\bf K}^2}{r+D_m{\bf K}^2}= 
        \Sigma({\bf K},0)\\
	z_2&\simeq&-(r+D_m{\bf K}^2).
	\label{z_2}
\end{eqnarray}
where the diffusion kernel is
\begin{equation}
	\Sigma({\bf K},0)=-\frac{gD_m{\bf K}^2}{r+D_m{\bf K}^2}=
g\left(\frac{1}{1+(\lambda{\bf K})^2}-1\right).
	\label{sigma_K}
\end{equation}
Neither the kernel nor the diffusivity now do not depend on $z$ and in
equations below this argument will be dropped.

The Laplace transform in Eq.\ (\ref{zz1z2}) reduces to the calculation
of pole residues:
\begin{eqnarray}
	&&G({\bf K},t)=e^{tz_1}
+\frac{z_1z_2}{z_1-z_2}\left(e^{tz_2}-e^{tz_1}\right)\nonumber\\
&&\simeq e^{t\Sigma({\bf K})}
+{\cal D}({\bf K}){\bf K}^2\left(e^{tz_2}
	-e^{t\Sigma({\bf K})}\right).
	\label{GKt}
\end{eqnarray}
As can be seen, in real space the second term on the second line would
integrate to zero while the first one to unity because the spacial
integration corresponds to the Fourier component ${\bf K=0}$. This
separates Eq.\ (\ref{GKt}) into contributions of different order in
$c_v$ as follows. Following Cowern et al.,\cite{cowern1990,cowern1991}
let us consider the long-time diffusion when the number of impurities
from the initial profile that experienced one or more encounters with the
vacancies is comparable to the number of all impurities in the profile,
i.\ e., is of order $O(1)$.  This means that $gt=O(1)$ which in its
turn means $t=O(1/c_v)$.  Now because ${\cal D}=O(c_v)$, the products
$t{\cal D}$ in the first and in the last exponential functions are of
order unity. But the factor ${\cal D}$ before the second term makes
the contribution due to the last exponential function negligible if we
are interested only in $O(1)$ terms. The only potentially problematic
term is the first exponential function in the parentheses that contains
$t=O(c_v^{-1})$ without compensating $O(c_v)$ factor because $z_2=O(1)$
as can be seen from Eq.\ (\ref{z_2}). But as is easy to see,
\begin{equation}
	e^{tz_2}\leq e^{-rt}\ll 1
	\label{inequalities}
\end{equation}
because $r=O(1)$ while $t=O(c_v^{-1})$. Thus, the second term in Eq.\
(\ref{GKt}) is much smaller than the first term and can be neglected if
we agree to neglect $O(c_v)$ contributions.

Thus, the leading $O(1)$ terms in Eq.\ (\ref{GKt}) can be reduced to the 
following Fourier-transformed diffusion equation 
\begin{equation}
\frac{\partial C({\bf K},t)}{\partial t}=
\Sigma({\bf K})C({\bf K},t),
	\label{dif-eqK}
\end{equation}
where $C$ is the Fourier transform of impurity concentration.  In real
space the diffusion equation Eq.\ (\ref{dif-eqK}) turns out to be not a 
differential equation but an integro-differential one
\begin{equation}
	\frac{\partial C({\bf R},t)}{\partial t}=\int 
	\Sigma({ \bf R-R}_0)C({\bf R}_0,t)d{ \bf R}_0
	\label{dif-eqR}
\end{equation}
with the diffusion kernel
\begin{equation}
	\Sigma({ \bf R-R}_0)=gG^P({ \bf R-R}_0)-g\delta({ \bf R-R}_0).
	\label{kernel}
\end{equation}
Here $G^P$ is the GF of the screened Poisson's equation from Eq.\
(\ref{poisson}) and thus describes the limiting ($t\to\infty$)
profile of the pair diffusion. This makes transparent the physical
meaning of Eq.\ (\ref{dif-eqR}). The impurity at point ${\bf R}_0$
is picked up by a vacancy with the rate $g$ (the second term on the
r.h.s.) and via the pair diffusion is redistributed with the probability
density $G^P$ (the first term on the r.h.s.).  The time dependence
of this process is ignored because Eq.\ (\ref{dif-eqR}) describes
diffusion on the time scale $O(c_v^{-1})$ which is much larger than
the $O(r^{-1})=O(1)$ scale of the pair diffusion.  This is exactly the
physics studied in Refs.\ \onlinecite{cowern1990,cowern1991}. From
Eq.\ (\ref{dif-eqR}) is easy to understand the non-Fickian
character of the pair-mediated diffusion discussed in Refs.\
\onlinecite{cowern1990,mirabella_mechanisms_2013}. Imagine a host with
inhomogeneous distribution of impurities in it and a bounded region
within that is completely devoid of them.  Despite this, the rate of
growth of the impurity concentration inside the region will be everywhere
positive according to Eq.\ (\ref{dif-eqR}) because of the spatially
extended diffusion kernel Eq.\ (\ref{kernel}) that is able to displace
impurities at finite distances. This starkly contrasts with the local
current picture underlying the Fickian diffusion.

\begin{figure}
\begin{center}
\includegraphics[viewport = 0 0 400 370, scale = 0.6]{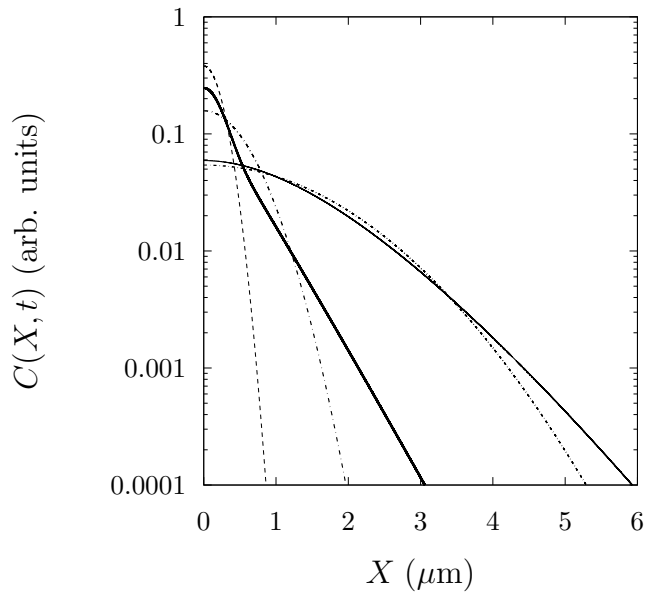}
\end{center}
\caption{\label{50Cprofiles} 1D profiles for the diffusion of lanthanum
impurity in aluminum at temperature $50\,^\circ\mbox{C}$ starting from
the initial Gaussian distribution of with 0.5~$\mu m$ (dashed line).
Thick solid line: the profile calculated according to Eq.\ (\ref{C(X,t)})
for $t=5$~hours; thin solid line: $t=50$~hours; dashed-dotted lines: the
profiles at 5 and 50~hours as predicted by the conventional diffusion
equation; the time intervals were chosen to correspond to $gt\simeq1$
and $\simeq10$, respectively.}
\end{figure}

3D diffusion equation Eq.\ (\ref{dif-eqR}) makes possible numerical
simulation of the diffusion profiles in any geometry.  In large systems
the solution should presumably be sought via its direct numerical
integration. In 3D the Poisson kernel is singular (it coincides with
the screened Coulomb or the Yukawa potential) but efficient techniques
of dealing with it were proposed in Ref.\ \onlinecite{poisson_eq}. In
systems of moderate sizes a convenient method provides the Fourier
transform. In 1D case the solution reads
\begin{equation}
	C(X,t)=\frac{1}{2\pi}\int_{-\infty}^\infty e^{t\Sigma(K)}C_0(K)dK,
	\label{C(X,t)}
\end{equation} 
where $C_0$ is the Fourier transform of the initial profile. In this way
were calculated NGDPs presented in Fig.\ \ref{50Cprofiles} where $C_0$
where chosen to be Gaussian to easier visualize the NGDPs caused by
the pair-mediated diffusion (in conventional diffusion an initially
Gaussian-shaped profile remain Gaussian at all times).  As shown in
Appendix \ref{app}, the 1D profiles calculated within our approach
coincide with those of Refs.\ \onlinecite{cowern1990,cowern1991}
so in Fig.\ \ref{50Cprofiles} one can see the exponential tails in
the diffusion profiles similar to those found by Cowern et al. The
tails, however, cannot extend on arbitrarily long distances because
the diffusion kernel in Eq.\ (\ref{sigma_K}) at small Fourier momenta
behaves as $\sim{\bf K}^2$ so from the inverse Fourier transform of the
kind of Eq.\ (\ref{C(X,t)}) but for arbitrary dimension it is easy to see
that at finite times the large-$|{\bf R}|$ asymptotic will be Gaussian,
similar to the case of diffusion of individual pair discussed in Sec.\
\ref{unstable-diff}.
\section{\label{conclusion}Conclusion}
In the present paper a theory of the vacancy-mediated diffusion
in the case of strong I-v binding has been developed. It has
been shown that tightly bound I-v pairs provide the mobile state
of impurities that underly the NGDPs similar to those observed
in dopant diffusion in semiconductors and in the copper surface
layers.\cite{cowern1990,cowern1991,mirabella_mechanisms_2013,%
In-V-attraction,In/cu(001)} By unifying the phenomenological theory
of Cowern et al.\cite{cowern1990} with the 5FM of the vacancy-mediated
diffusion in FCC hosts\cite{lidiard,LECLAIRE1978} it has been possible
to calculate numerical values of the parameters of the phenomenological
theory on the basis of the available data on the parameters of the
5FM.\cite{manning,philibert,voglAlFe,PhysRevB.43.9487,LDA+UinAl,%
Wolverton2007,ab_init_Al2009,point_defects_ab_initio2014,%
wu_high-throughput_2016,Wu2017} This has made possible identification
of the impurity-host systems suitable for the observation of the NGDPs
as well as their explicit simulation in La\underline{Al} system where
the phenomenon is expected to be the most pronounced among the solutes
in aluminum host. Because the NGDPs are universal,\cite{cowern1990} all
impurities listed in Table \ref{table2} should exhibit the same profiles
as shown in Fig.\ \ref{50Cprofiles} but at shorter length scales. The
latter can be enlarged by lowering the temperature but the time of the
observation will have to be extended correspondingly.

Apart from the calculation of the parameters of the phenomenological
model in the framework of the 5FM, the approach of Cowern et
al.\cite{cowern1990} has been extended in two respects. First,
it has been shown that the diffusion mode studied in Refs.\
\onlinecite{In-V-attraction,In/cu(001)} in 2D can contribute to 3D
NGDPs in cases when the initial state already contains associated I-v
pairs.  Because their diffusion starts immediately, they introduce
impurity diffusion on much shorter time scale than that of Ref.\
\onlinecite{cowern1990}.  For example, in La\underline{Al} system at
room temperature (20$^\circ$~C) the parameter $\lambda\simeq2~\mu m$.
The conventional diffusion will cower this distance in over a month
while the pair diffusion in about three hours. Accounting for this
mode of diffusion may be of practical importance in assessment of
the longevity of microelectronic devices. Though the concentration
of preexisting pairs is usually expected to be small, in some of the
projected devices the functional elements will consist of only one
atom\cite{kane_silicon-based_1998} so the estimates of the longevity that
neglect the fast diffusion of the small number of contaminating atoms
that are associated with the vacancies introduced during the deposition
process may lead to serious errors.

The second extension of the theory of Ref.\ \onlinecite{cowern1990} has
been achieved through its blending with the Dyson equation.  This resulted
in a non-Fickian integro-differential diffusion equation describing the
pair-propagated impurity diffusion in arbitrary geometry that can be
used in simulations of NGDPs in any elemental hosts.

Special attention in the paper has been devoted to the systems with
large $w_4/w_0$ ratio for two reasons.  First, the ratio turned out to
be large for all impurities in aluminum host with the largest values of
$\lambda$, i.\ e., in the systems that should be the most appropriate
for experimental study of NGDPs. Secondly, and more importantly,
this case makes possible accurate quantitative predictions about the
phenomenon. A serious problem of the microscopic diffusion theory is
that both experimental definitions and first-principle calculations
of various activation energies contain errors of order $O(0.1$~eV)
in the best case.\cite{ab_init_Al2009,wu_high-throughput_2016,Wu2017}
The jump frequencies and diffusion constants depend on the energies
via the Arrhenius law and at temperatures in a few hundred Kelvins
may be orders of magnitude off from their true values. But it has
been shown in the present paper that when $w_4/w_0$ is large the
phenomenological parameters $g$ and $\lambda$ can be calculated from
only two quantities: the impurity diffusion constant and that of
the host self-diffusion.  Both can be measured at the experimental
temperature independently and the two parameters calculated on
their basis can subsequently be used in the profile simulations.
The simulated NGDPs should agree with experimental ones quantitatively,
provided the 5FM is an adequate model for the system under study.
Significant discrepancies will mean that the canonical 5FM is too
simplistic for the case under consideration and needs to be improved
along the lines suggested by first-principles calculations and physical
considerations.\cite{LECLAIRE1978,Bocquet,wu_high-throughput_2016} This
conclusion relies on the assumption that the errors in in the frequencies
$w_4$ and $w_0$ are not too large to reduce our estimates of $w_4/w_0$
ratios for the systems listed in Table \ref{table1} more than 4--6
orders of magnitude.  Unfortunately, at present this possibility cannot
be completely excluded.

The important question that has not be adequately addressed in the
present paper concerns the reliability and the accuracy of the developed
theory which has been substantiated mainly by qualitative arguments
and phenomenological approaches.  This difficult question will be
addressed in a separate paper where a rigorous treatment of the 5FM in
the general case of the I-v interaction of arbitrary strength will be
presented.\cite{2follow} It will be shown that in the limit of strong I-v
attraction the results of the present paper are in excellent agreement
with the rigorous solution.\cite{cond-mat/0505019}
\begin{acknowledgments}
I would like to express my gratitude to Hugues Dreyss\'e for encouragement.
\end{acknowledgments}
\appendix
\section{\label{app}Comparison with NGDP of Ref.\ \onlinecite{cowern1990}}
In slightly modified notation, the 1D diffusion profile with initial
delta-function distribution was shown to be described by the series
given by Eqs.\ (5)-(10) of Ref.\ \onlinecite{cowern1990} as
\begin{equation}
C(\xi,\theta)=\lambda^{-1}\sum_{n=0}^\infty P_n(\theta)\phi_n(\xi,1),
	\label{C(xi)}
\end{equation}
where
\begin{equation}
	\theta = gt,\qquad \xi = X/\lambda,
	\label{xi-theta}
\end{equation}
\begin{equation}
	P_n(\theta)=(\theta^n/n!)e^{-\theta},
	\label{Pn}
\end{equation}
\begin{equation}
	\phi_{n=0}(\xi,1)=\delta(\xi),
	\label{fn=0}
\end{equation}
and
\begin{equation}
\phi_{n>0}(\xi,\mu)=\frac{e^{-\sqrt{\mu}|\xi|}}{(2\sqrt{\mu})^{2n-1}}
\sum_{k=0}^{n-1}\frac{2^k}{k!}\binom{2n-2-k}{n-1}(|\xi|\sqrt{\mu})^k.
\label{fn>0}
\end{equation}
Here we introduced the parameter $\mu$ to facilitate the proof that
the profile Eq.\ (\ref{C(xi)}) from Ref.\ \onlinecite{cowern1990}
coincides with the 1D profile from Eq.\ (\ref{C(X,t)}) with $C_0(K)=1$
for the delta-function initial profile:
\begin{eqnarray}
C(X,t)&=&\frac{1}{2\pi}\int_{-\infty}^\infty dK e^{iXK} e^{-gt}\exp\left(\frac{gt}{1+(\lambda K)^2}\right)\nonumber\\
&=&\frac{1}{\lambda}
\sum_{n=0}^\infty P_n(\theta) \frac{1}{2\pi}\int_{-\infty}^\infty 
\frac{e^{i\xi u}du}{(1+ u^2)^n}.
	\label{C(X)}
\end{eqnarray}
Here the last exponential on the first line has been expanded in the
Tailor series so by comparison with Eq.\ (\ref{C(xi)}) we conclude that
the inverse Fourier transforms on the second line should be equal to
$\phi_{n>0}(\xi,1)$. To show this we first introduce the integrals
\begin{equation}
	\phi_{n}(\xi,\mu)=\frac{1}{2\pi}\int_{-\infty}^\infty 
	\frac{e^{i\xi u}du}{(\mu+ u^2)^n}
	\label{phi(mu)}
\end{equation}
and note that if $\phi_{n=1}(\xi,\mu)$ is known, other integrals
can be computed recursively as
\begin{equation}
	\phi_{n+1}(\xi,\mu)=-\frac{1}{n}\frac{d}{d\mu}\phi_{n}(\xi,\mu).
	\label{n+1}
\end{equation}
Thus, we only need to show that $\phi_{n}(\xi,\mu)$ in Eq.\ 
(\ref{fn>0}) satisfy the recursion.  To this end we first note that
with the exponential factor being common to all terms in all functions,
the equality in Eq.\ (\ref{n+1}) will hold if it will be valid for every
power of $|\xi|^k$ under the summation sign. Let us consider one such
term in Eq.\ (\ref{fn>0})
\begin{equation}
	\phi_n^{(k)}=\frac{e^{-\sqrt{\mu}|\xi|}}{2^{2n-1}(n-1)!}
	\frac{2^k}{k!}\frac{(2n-2-k)!}{(n-1-k)!}|\xi|^k\mu^{(k+1)/2-n}.
	\label{phink}
\end{equation}
When substituted in Eq.\ (\ref{n+1}) it will contribute to $|\xi|^k$
term in $\phi_{n+1}^{(k)}$ through the derivative of its last factor with
respect to $\mu$. The only other contribution from $\phi_n$ contributing
into $|\xi|^k$ term in $\phi_{n+1}$ is $\phi_n^{(k-1)}$ differentiated
with respect to $\mu$ in the exponential function. It is straightforward
to check that these two contributions lead to the term $\phi_{n+1}^{(k)}$
as in Eq.\ (\ref{phink}) only with $n+1$ instead of $n$, as required.
\end{document}